\documentclass[a4paper]{PoS}
\usepackage{ mathrsfs }
\usepackage{slashed}
\usepackage{amsmath}
\title{Strangeness production in heavy-ion collisions}

\ShortTitle{Strangeness production in heavy-ion collisions}

\author{\speaker{Alessia Palmese}%
\\ Institut f{\"u}r Theoretische Physik, Universit{\"a}t Gie{\ss}en, 35392 Gie{\ss}en, Germany
 \\       E-mail: \email{alessia.palmese@theo.physik.uni-giessen.de}}

\author{Giuseppe Pagliara, Alessandro Drago\\
Dipartimento di Fisica - Universit{\`a} di Ferrara and INFN Sez. di Ferrara, 44122 Ferrara,Italy
}
\author{Olena Linnyk, Wolfgang Cassing\\
Institut f{\"u}r Theoretische Physik, Universit{\"a}t Gie{\ss}en, 35392 Gie{\ss}en, Germany
}

\abstract{
A study of the "horn" in the particle ratio $K^+/\pi^+$ for central heavy-ion collisions as a function of the collision energy $\sqrt{s}$ is presented. We analyse two different interpretations: the onset of deconfinement and the transition from a baryon- to a meson-dominated hadron gas. We use a realistic equation of state (EOS), which includes both hadron and quark degrees-of-freedom. The Taub-adiabate procedure is followed to determine the system at the early stage. Our results do not support an explanation of the horn as due to the onset of deconfinement. Using only hadronic EOS we reproduced the energy dependence of the $K^+/\pi^+$ and $\Lambda/\pi^-$ ratios employing an experimental parametrisation of the freeze-out curve. We observe a transition between a baryon- and a meson-dominated regime; however, the reproduction of the $K^+/\pi^+$ and $\Lambda/\pi^-$ ratios as a function of $\sqrt{s}$ is not completely satisfying. We finally propose a new idea for the interpretation of the data, the roll-over scheme, in which the scalar meson field $\sigma$ has not reached the thermal equilibrium at freeze-out.
The rool-over scheme for the equilibration of the $\sigma$-field is based on the inflation mechanism.
The non-equilibrium evolution of the scalar field influences the particle production, e.g. $K^+/\pi^+$, however, the fixing of the free parameters in this model is still an open issue. 
}

\FullConference{9th International Workshop on Critical Point and Onset of
Deconfinement\\
                 17-21 November, 2014\\
                 ZiF (Center of Interdisciplinary Research), University of Bielefeld, Germany}

\begin{document}
\section{Introduction}
High energy collisions are the unique experimental way to study the Equation of State (EOS) of strongly interacting matter.
New experiments on this topic are currently in preparation such as the CBM at FAIR and the MPD at NICA. In order to extract information about the behaviour of matter, one has to analyse the data with respect to the production of various hadronic particles, both mesons and baryons. This analysis is highly non-trivial, since at these energy regimes the baryon chemical potential of matter cannot be neglected and, unfortunately, the EOS of the matter at finite baryon chemical potential is still uncertain.
The possibility to investigate the deconfinement phase transition at high density and temperature is one of the main goals of the scienific community in the recent years. 
Assuming that the system at the early-stage of the heavy-ion collisions enters the Quark-Gluon-Plasma (QGP) phase, during the expansion of the fireball the system cools down so that the produced strong matter hadronizes into confined states.  A transition of this kind, from QGP phase to hadron phase, probably occurred during the expansion of the early universe about one microsecond after the Big Bang. 
But, even if the system reaches the deconfined phase, the life time of the fireball is extremely short ($\sim10^{-22}seconds$).
Among the suggested  experimental signatures of the QGP formation, the production of strange particles is one of the most investigated.
Indeed, in the data on the collision energy dependence of the ratio between the $K^+$ and the $\pi^+$ mesons a "horn" structure appears at  $\sqrt{s_{NN}}$ of about $7\,GeV$, whose origin is still unclear.
In this contribution, we aim at reanalysing the two main scenarios for the explanation of the "horn" \cite{Gazdzicki:1998vd}, \cite{Andronic:2009jd} using an up-to-date EOS. In addition, we propose a new idea to describe particle production in heavy-ion collisions.
In section $2$ we describe the EOS used for the calculations; in section $3$ we present the main features of the three models, while in section $4$ we show our results achieved for each approach. In section $5$ we draw our conclusions.

\section{Equation of state}

A reliable modeling of the equation of state is essential to describe the system produced in heavy-ion collisions. 
In this section we are going to draw only the main features of the EOS used; for a detailed description we address the reader to reference \cite{Lavagno:2010ah}.
The considered EOS includes hadron degrees-of-freedom, a hadron-quark mixed phase and quark degrees-of-freedom at small, intermediate and high baryon densities, respectively. The system is characterized by 3 conserved charges: baryon number, electric charge Z/A = 0.4, zero strangeness. 
The hadronic part of the EOS is built according to the relativistic mean-field model based on a Walecka Lagrangian, with the inclusion of the baryon octet and of the $\Delta$-isobars.
The interactions between hadrons are included via meson exchange whose mediators are $\sigma$, $\pi$, $\omega$, $\rho$ meson fields. 
The values of the mass, energy and chemical potential of each baryon are modified by the interaction.
The Walecka model in the relativistic mean-field approximation allows to describe the properties of finite nuclei and of dense and finite-temperature nuclear matter \cite{Muller:1995ji}. 
We include in the model also the lightest pseudo-scalar and vector mesons.
We take into account the one-body meson contribution simply by considering the mesons as an ideal Bose gas with effective chemical potentials, which ensures the self-consistent interaction of the mesons. The introduction of the effective chemical potential for the mesons can be compared with the excluded volume approximation of the hadron resonance gas \cite{Satarov:2009zx}.
In the deconfined state the EOS is given by the MIT bag model.  
We include the $u$, $d$ and $s$ quarks as relativistic Fermi particles.
The $u$ and $d$ quarks are assumed massless, while the $s$-quark mass is finite and is fixed to $95MeV$.
The gluons are considered as massless Bose particles with zero chemical potential.
Finally, the mixed phase is realized by imposing the Gibbs conditions for a system with three conserved charges, which are baryon number, electric charge and strangeness number.
We mention that our equation of state does not satisfy chiral symmetry.

\section{Models}

\subsection{Statistical Model of the Early Stage (SMES)}
The Statistical Model of the Early Stage (SMES) has been introduced by Gazdzicki and Gorenstein \cite{Gazdzicki:1998vd}, \cite{Gazdzicki:2010iv}. It is a statistical model describing the heavy-ion collision process, which turns out to be in good agreement with experimental data for different hadron abundancies. This approach provides important information about the features of the system created at the early stage of the collision and is able to suggest signatures for the onset of QGP formation. The SMES proposes that the sharp maximum of $K^+/\pi^+$ at about $\sqrt{s_{NN}}\approx 7.5GeV$ corresponds to the onset of deconfinement. Moreover, within this approach, the strangeness to entropy ratio $(n_s+n_{\bar s})/s$  is strictly connected to the multiplicity ratio $<K^+>/<\pi^+>$ and is characterised by an analogous non-monotonic behaviour.
Finally, according to the SMES, the mixed phase is reached for the beam energies of $30A-60A GeV$ ($\sqrt{s_{NN}}=7-12 GeV$).
The SMES deals with a system with vanishing chemical potentials and baryon density, besides most degrees-of-freedom are taken to be massless.
The first assumption relies on the hypothesis that after the collision process the wounded nuclei go away from the collision point with velocity close to $c$, leaving a system with vanishing baryon density. It is extremely difficult to evaluate the initial conditions of the evolving system, but recent works suggest that the early system has a non-negligible baryon density \cite{Arsene:2006vf}, \cite{Merdeev:2011bz}. Concerning the second assumption, the SMES seems to give realistic masses only to the strange degrees-of-freedom, but, especially at low energies, the temperatures reached by the system in the confined state are not large enough to neglect the masses of the lightest non-strange particles. 

\subsection{Hadron Resonance Gas Models}
Several Hadron Resonance Gas Models have been developed to explain the particle production in heavy-ion physics, achieving good results also for the interpretation of the horn (\cite{Bugaev:2014cha}, \cite{Merdeev:2011bz} and \cite{Andronic:2009jd}).
The particle production in heavy-ion collision is described as a thermal hadron production at the freeze-out stage. The freeze-out represents the final stage of the expansion process, in which the produced hadrons cease to interact and, as a result, the particle multiplicities are frozen. These models do not propose a description of the system evolution and they do not need any information about the initial stage of the collision. The matter produced in the collisions is represented as a Hadron Resonance Gas, in which all hadrons containing $u$, $d$ and $s$ quarks are included as free particles usually up to a mass of about $2-3\,GeV$. The energy dependence of the particle ratios, e.g. of $K^+/\pi^+$ and $\Lambda/\pi^-$, is provided computing the particle densities along the paremetrisation of the experimental freeze-out line, fitted to reproduce the yields of non-strange hadrons. In order to correctly reproduce the data, some models, e.g. the Statistical Hadronization Model (SHM) \cite{Becattini:2003wp}, introduce non-equilibrium parameters to modify the expressions of the particle densities. 
In our work we follow the approach of Ref. \cite{Cleymans:2004hj} and \cite{Andronic:2009jd}. 
These Hadron Resonance Gas models predict a transition of the hadron gas from a baryon-dominated to a meson-dominated gas at $\sqrt{s_{NN}}\sim8\,GeV$ along the freeze-out curve. The origin of the horn is related to this type of transition. 

\subsection{Roll Over scheme}
Here, we describe a possible mechanism which can give information about the chemical equilibrium of the system during its expansion after the stage of maximum compression. If the system is not in full chemical equilibrium, the particle densities and the strangeness production can be affected by large fluctuations. Thus we propose an outlook for the study of the strangeness production in a partial equilibrium condition.  

\begin{figure}[h!]
\begin{minipage}[b]{0.5\textwidth}
\centering
\includegraphics[width=1.1\textwidth]{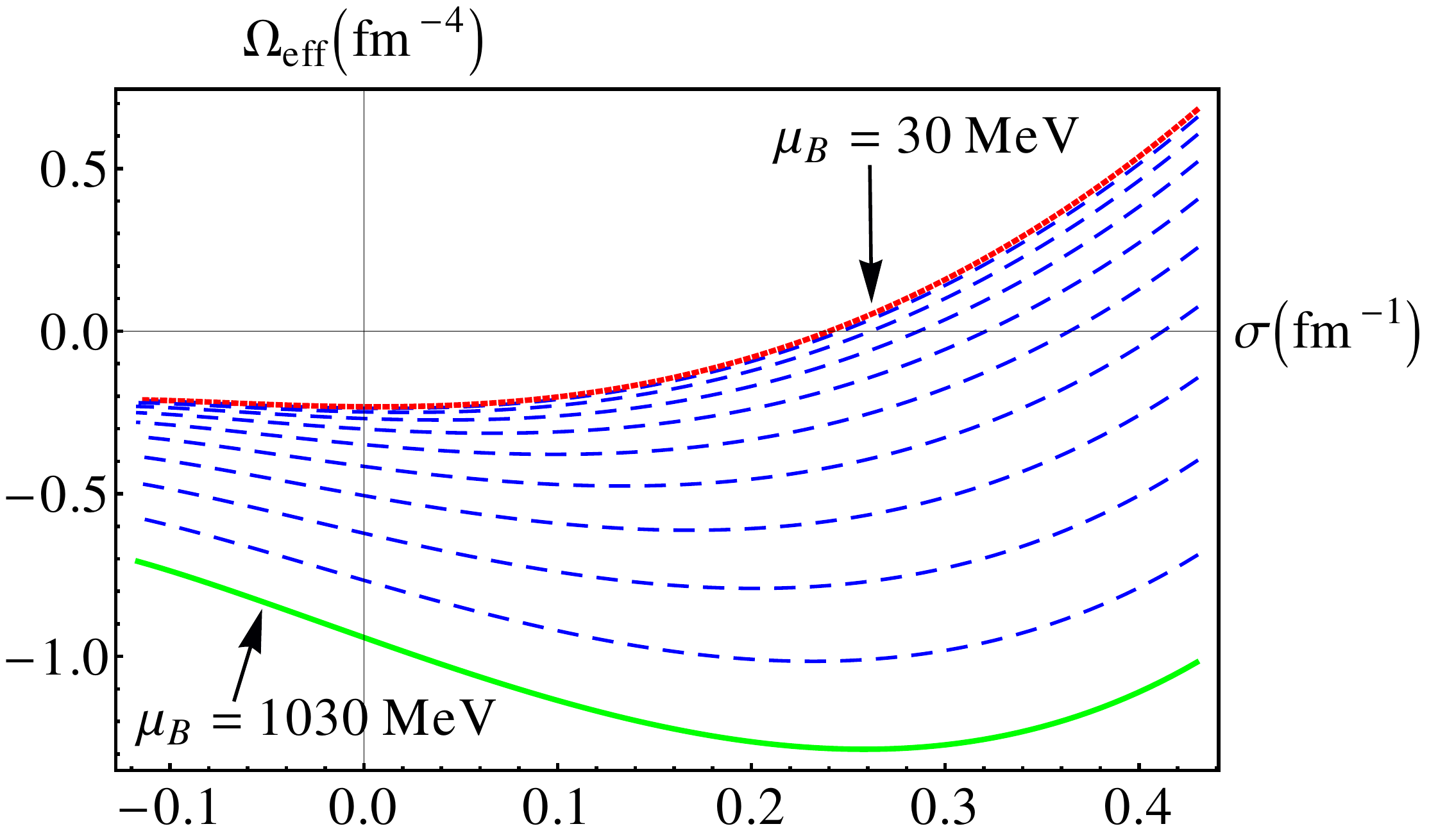}
\caption{$\Omega_{eff}(\sigma,T)$ as a function of $\sigma$ at $T=156\,MeV$; the green solid line corresponds to the upper value of baryon chemical potential $\mu_B=1030\,MeV$, the red dotted line corresponds to the lower value of baryon chemical potential $\mu_B=30\,MeV$, the blue dashed lines correspond to the values of $\mu_B$ between $30\,MeV$ and $1030\,MeV$ with steps of $100\,MeV$.}
\label{fig:potmu}
\end{minipage}
\begin{minipage}[b]{0.05\textwidth}
$ $

\end{minipage}
\begin{minipage}[b]{0.45\textwidth}
\centering
\includegraphics[width=1.1\textwidth]{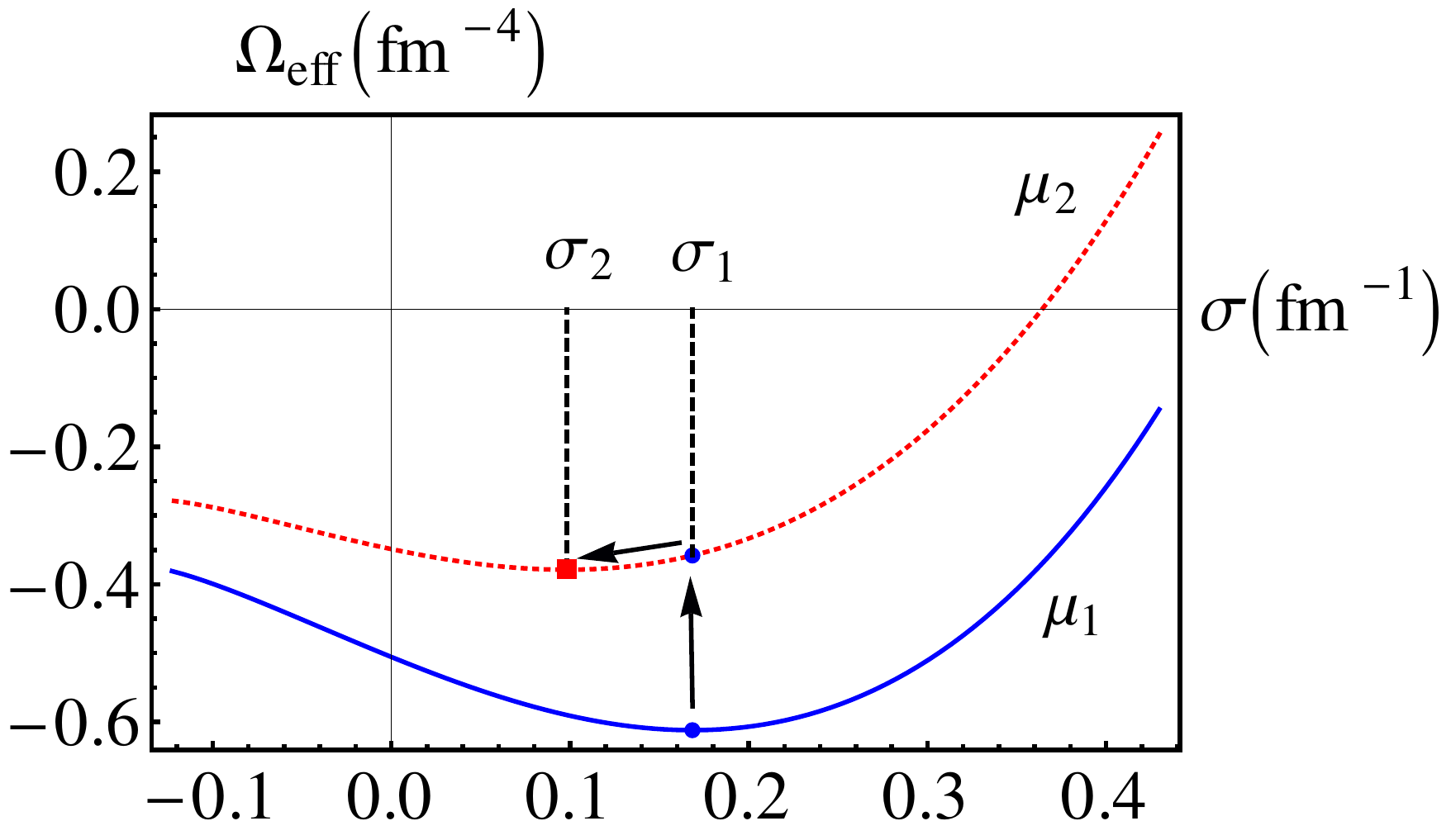}
\caption{Roll over picture; the blue-solid and red-dotted curves represent $\Omega_{eff}$ at $T=156\,MeV$ for $\mu_1=750\,MeV$ and $\mu_2=550\,MeV$ respectively; the blue-circle points refer to the value of the field $\sigma_1$ associated to the minimum of the potential for $\mu_1$, the red-square point $\sigma_2$ represents the minimum of the potential for $\mu_2$.}
\label{fig:potroll}
\end{minipage}
\end{figure}

The heavy-ion collision can be seen as a Little Bang, since the system undergoes a rapid expansion and it reaches values of the energy density, characteristics of the early Universe. We can proceed with this parallelism also for the description of the evolution of the fireball. We know that after the Big Bang the Universe started expanding and during its evolution it cooled down, similar to the fireball produced at the heavy-ion collision process. 
The inflation is the exponential expansion of the universe driven by the dynamical evolution of a scalar field $\phi$ \cite{roll}. At the origin of the Universe, this field was not in the minimum of its potential, consequently it began to roll over towards the equilibrium point. We borrow this pitcure and apply it to the scalar field $\sigma$ as an inflaton field.
We choose $\sigma$ for this picture not simply because it is a scalar field, instead of the other vector mediator fields  $\omega$ and $\rho$, but we also take into account the important role that it plays in the system. In fact, the $\sigma$ field modifies the baryon masses, consequently a change in the $\sigma$ field directly influences the particle composition of the system.
The equation of motion for the scalar field is taken as:
\begin{equation}
\ddot\sigma+3\frac{1}{\tau}\dot\sigma+\Omega_{eff}'(\sigma)=0,
\label{roll}
\end{equation}
where $1/\tau$ represents the "Hubble parameter" for the expansion of the fireball. We define the rate of the fireball expansion as $\dot V/V$ (with $V$ as the volume of the fireball), which is approximately $1/\tau$.
\\Since we are dealing with a system composed not only of the $\sigma$ field, the potential to be considered in the equation of motion is the grand-canonical potential $\Omega$ of the whole system. 
In addition, since the temperature of the system is finite, we have to replace the potential $\Omega(\sigma,T)$ with the effective potential $\Omega_{eff}(\sigma,T)$, obtained as the difference between the grand canonical potential $\Omega$ and the thermal contribution associated to the field $P_\sigma(\sigma,T)$ (see \cite{Drago:2001gd} for details):
\begin{equation}
\Omega_{eff}(\sigma,T)=\Omega(\sigma,T)-P_\sigma(\sigma,T).
\end{equation} 
In \figurename~\ref{fig:potmu} we show the behaviour of $\Omega_{eff}(\sigma,T)$ for different values of the baryon chemical potential at a fixed temperature $T=156\,MeV$.
We notice that at a given $\sigma$ the potential $\Omega_{eff}$ increases as the baryon chemical potential decreases, reaching positive values at large values of $\sigma$ and for small $\mu_B$. Furthermore, the minimum of the potential shifts towards smaller values of $\sigma$ as the baryon chemical potential decreases. The curvature of the potential is larger for high $\mu_B$, whereas $\Omega_{eff}$ becomes almost flat for small $\mu_B$. As a result, the minimum of the potential is well-defined when the baryon chemical potential of the system is large, while for small $\mu_B$ the minimum of the potential is situated in a wide flat region. Thus, we can infer that in case of small $\mu_B$ the field can easily fall into non-equilibrium configurations. 
The effective potential is characterised by the same trend as a function of temperature: with decreasing temperature, the $\Omega_{eff}$ becomes larger and shallower and the minimum of the potential shifts to smaller values of the field. 
\\We know that during the collision the system heats up, reaching large values of $T$ and $\mu_B$ depending on the beam energy. During the expansion, the system cools down and the associated $\mu_B$ decreases (3-fluid hydrodynamics \cite{ivanov} is able to display how the system produced by a heavy-ion collision evolves in the $(T, \mu_B)$ plane).
In the present exploratory work, we study the expansion of a fireball produced in a heavy-ion collision at low energy $E_{Lab}=10AGeV$. The system does not undergo a deconfinement transition and during the expansion it evolves from a larger value of baryon chemical potential $\mu_1$ to a smaller value $\mu_2$ (\figurename~\ref{fig:potroll}). We consider in first approximation the temperature to be constant. 

\section{Results}
\subsection{SMES approach}
We use our equation of state to follow the main concepts of the SMES: the creation of new degrees-of-freedom occurs at the early stage as a statistical process; strange and antistrange particle abundancies are conserved during the expansion of the system.
It is very hard to estimate the initial conditions of the thermodynamical evolution of the system. In this work we choose to address this issue by following the procedure of the Taub-adiabate \cite{Merdeev:2011bz}.   
\begin{figure}[h!]
\begin{minipage}[b]{0.5\textwidth}
\centering
 \includegraphics[width=0.9\textwidth]{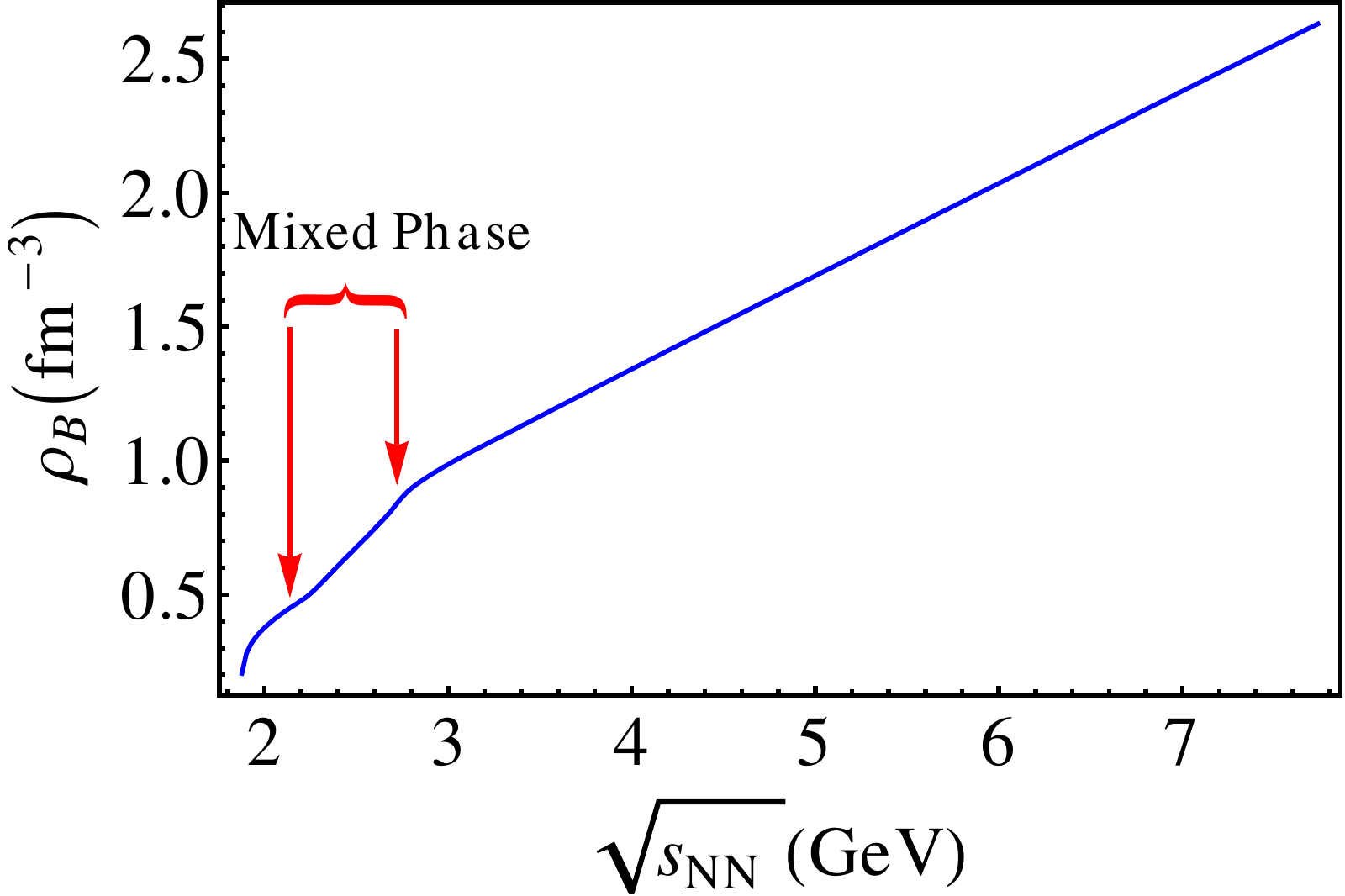}
\end{minipage}
\hfill
\begin{minipage}[b]{0.5\textwidth}
\centering
 \includegraphics[width=\textwidth]{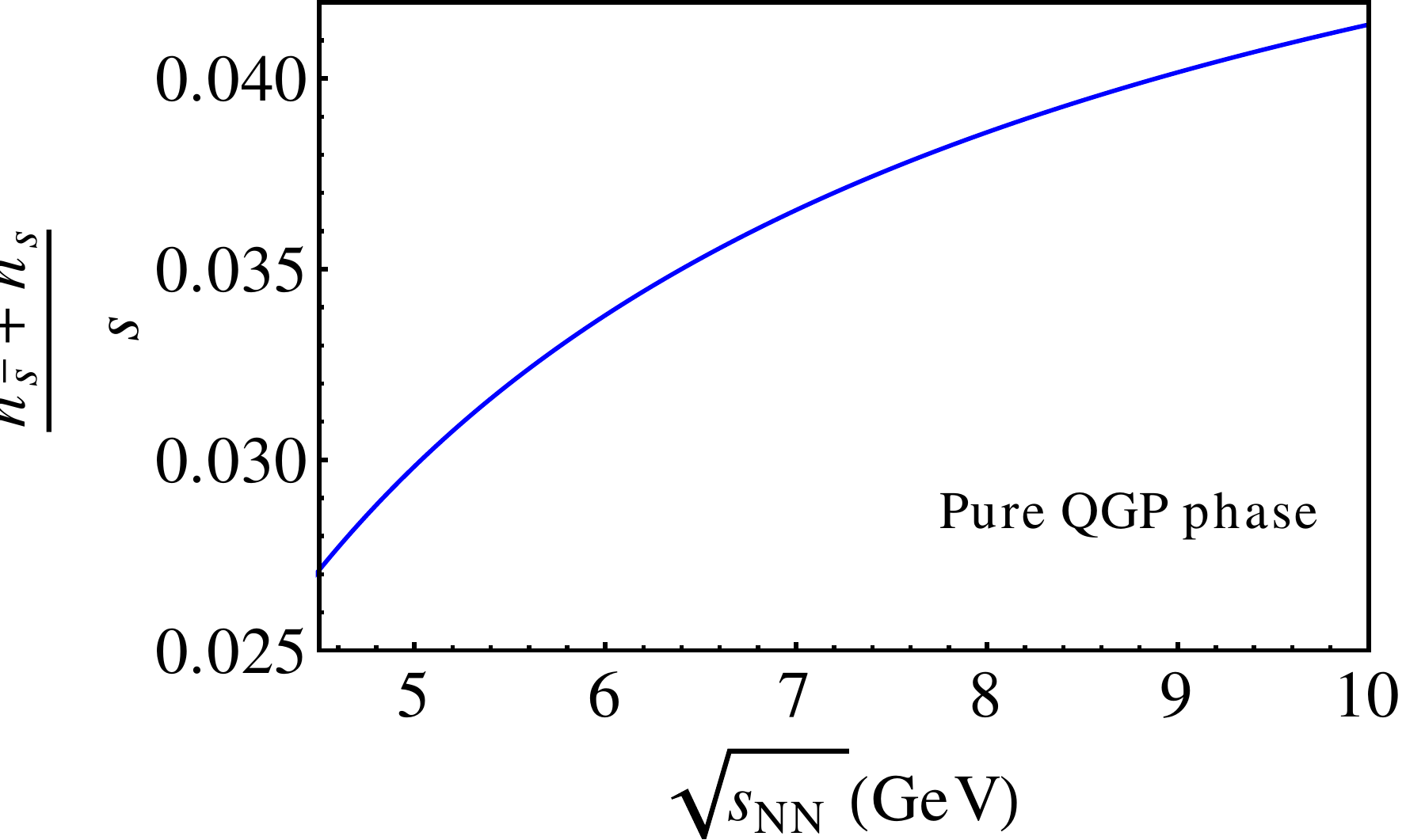}
\end{minipage}
\caption{Left panel: baryon density for the early stage system as a function of the center-of-mass energy; the arrows point out the kinks of $\rho_B$ related to the beginning and the end of the mixed phase. Right panel: the ratio $(n_s+\bar n_s)/s$ as a function of $\sqrt{s_{NN}}$ at the early stage in a pure QGP phase for our EOS.}
\label{fig:taub+ns}
\end{figure}
The baryon density, obtained through this procedure, is plotted in \figurename~\ref{fig:taub+ns} (left panel) as a function of $\sqrt{s_{NN}}$.
It exhibits two kinks when the system enters and leaves the mixed phase. The resulting values for the baryon chemical potential and baryon density are not small enough to be considered negligible, as supposed by the SMES. Moreover we can see that the mixed phase provided by our EOS ends at $\sqrt{s_{NN}}\sim3\,GeV$.
Consequently, it is not surprising that at $\sqrt{s_{NN}}\sim7\,GeV$ the strangeness to entropy ratio \figurename~\ref{fig:taub+ns} (right panel) does not show a peak associated to the onset of deconfinement. Anyway, the values of the critical densities are rather uncertain and  within our EOS the onset of deconfinement appears at quite small $\sqrt{s_{NN}}$, so we can not exclude that the SMES mechanism will work with other EOSs.

\subsection{Hadron Resonance Gas Models approach}
\begin{figure}[h!]
\begin{minipage}[b]{0.5\textwidth}
\centering
 \includegraphics[height=3.5cm]{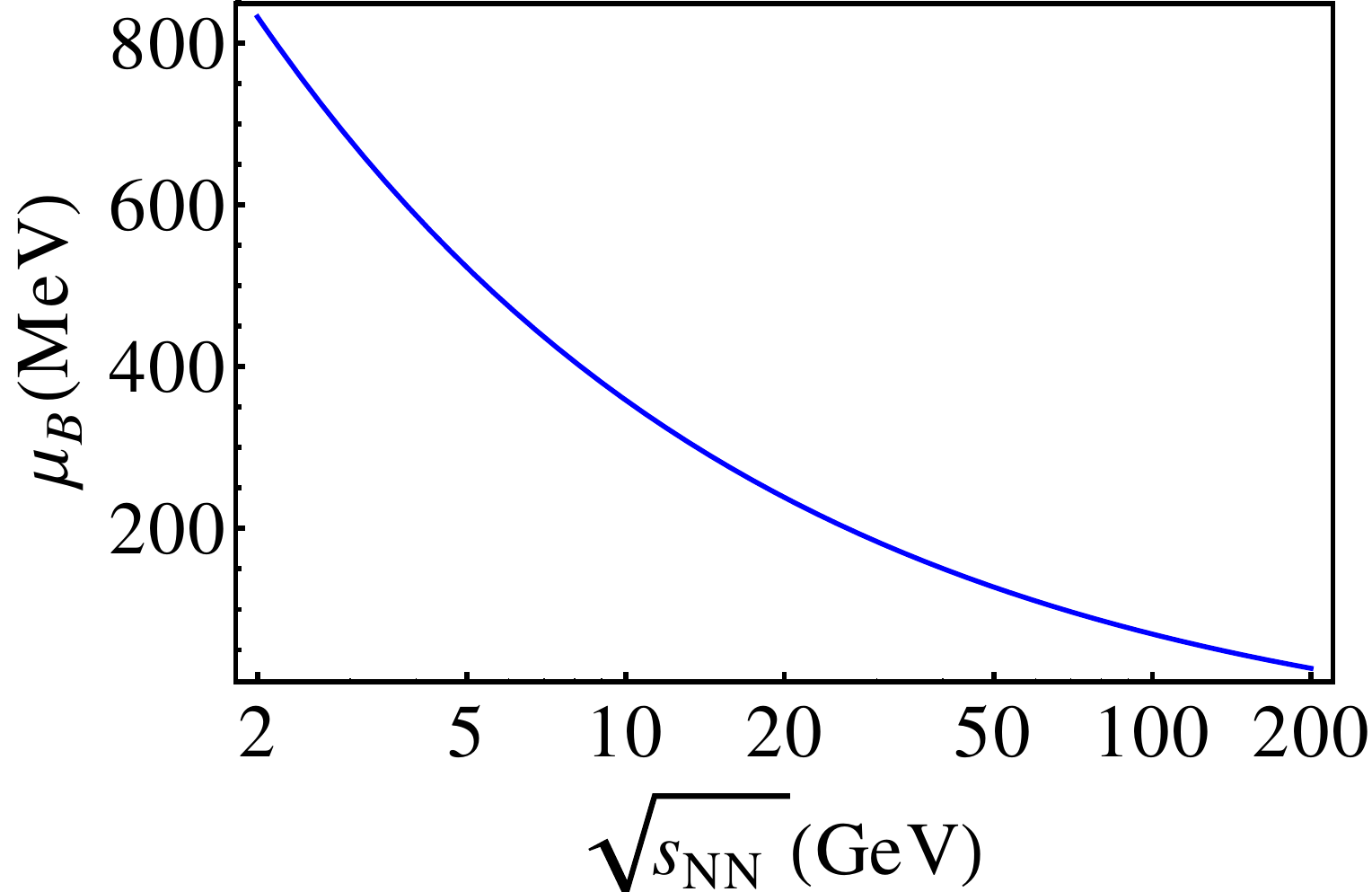}

\includegraphics[height=3.5cm]{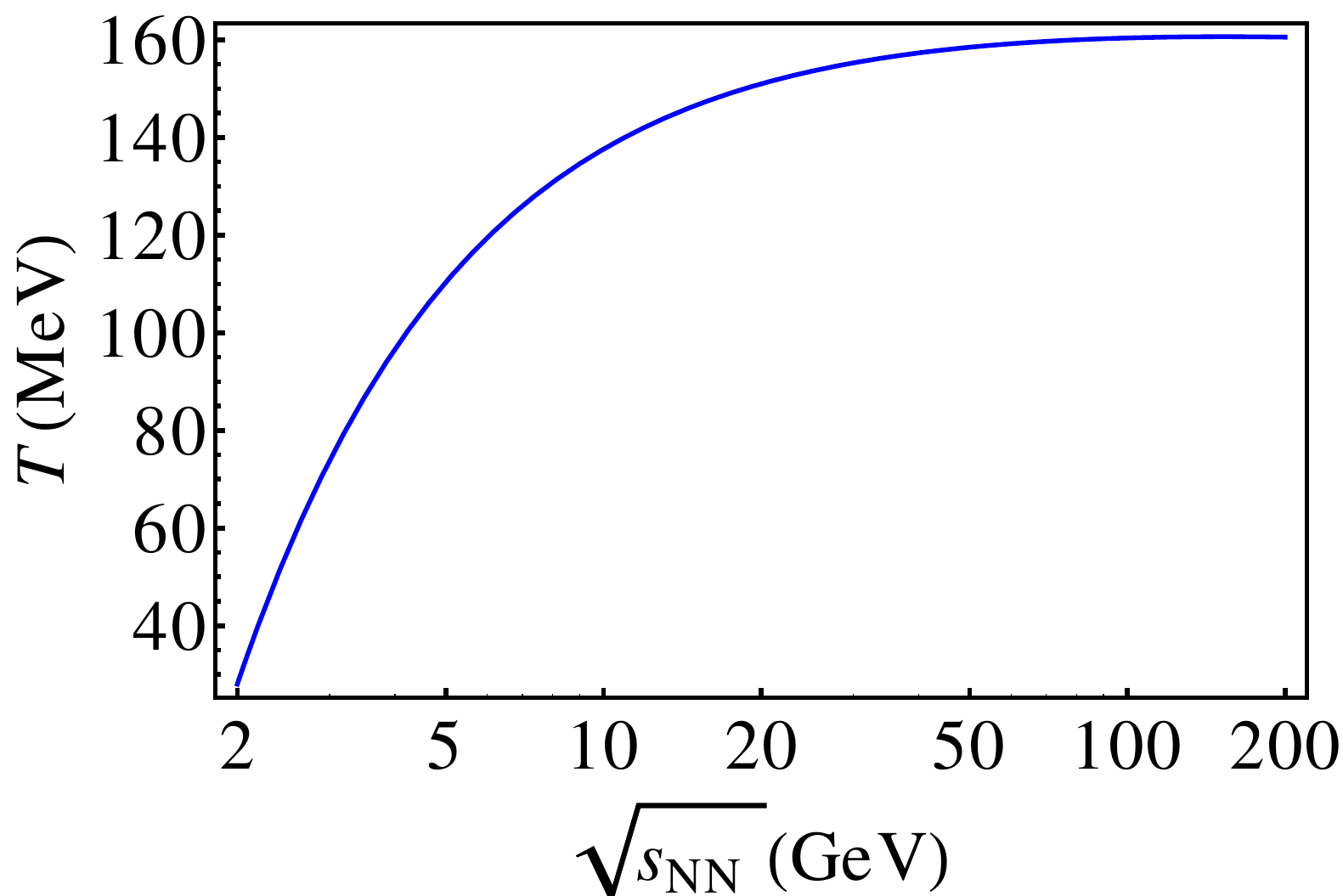}
\caption{Freeze-out parametrization.}
\label{fig:fos}
\end{minipage}
\begin{minipage}[b]{0.5\textwidth}
\centering
\includegraphics[width=\textwidth]{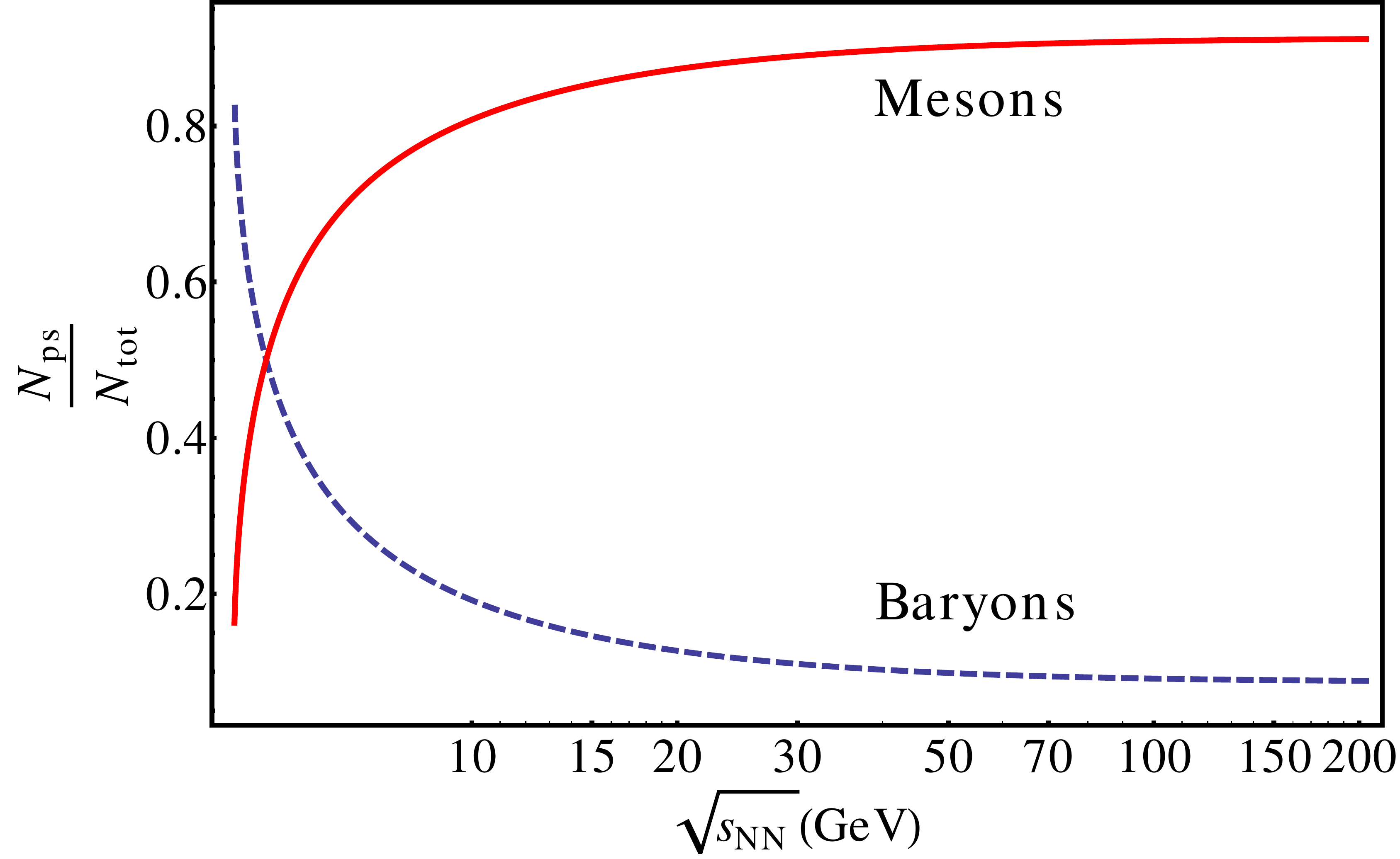}
\caption{The ratios between the particle species and the total particle densities as a function of $\sqrt{s_{NN}}$ (in logarithmic scale) at the freeze-out; the blue-dashed and red-solid lines refer to $n_B/n_{tot}$ and to $n_M/n_{tot}$ respectively.}
\label{fig:regime}
\end{minipage}
\end{figure}

\begin{figure}[h!]
\begin{minipage}[b]{0.5\textwidth}
\centering
\includegraphics[width=\textwidth]{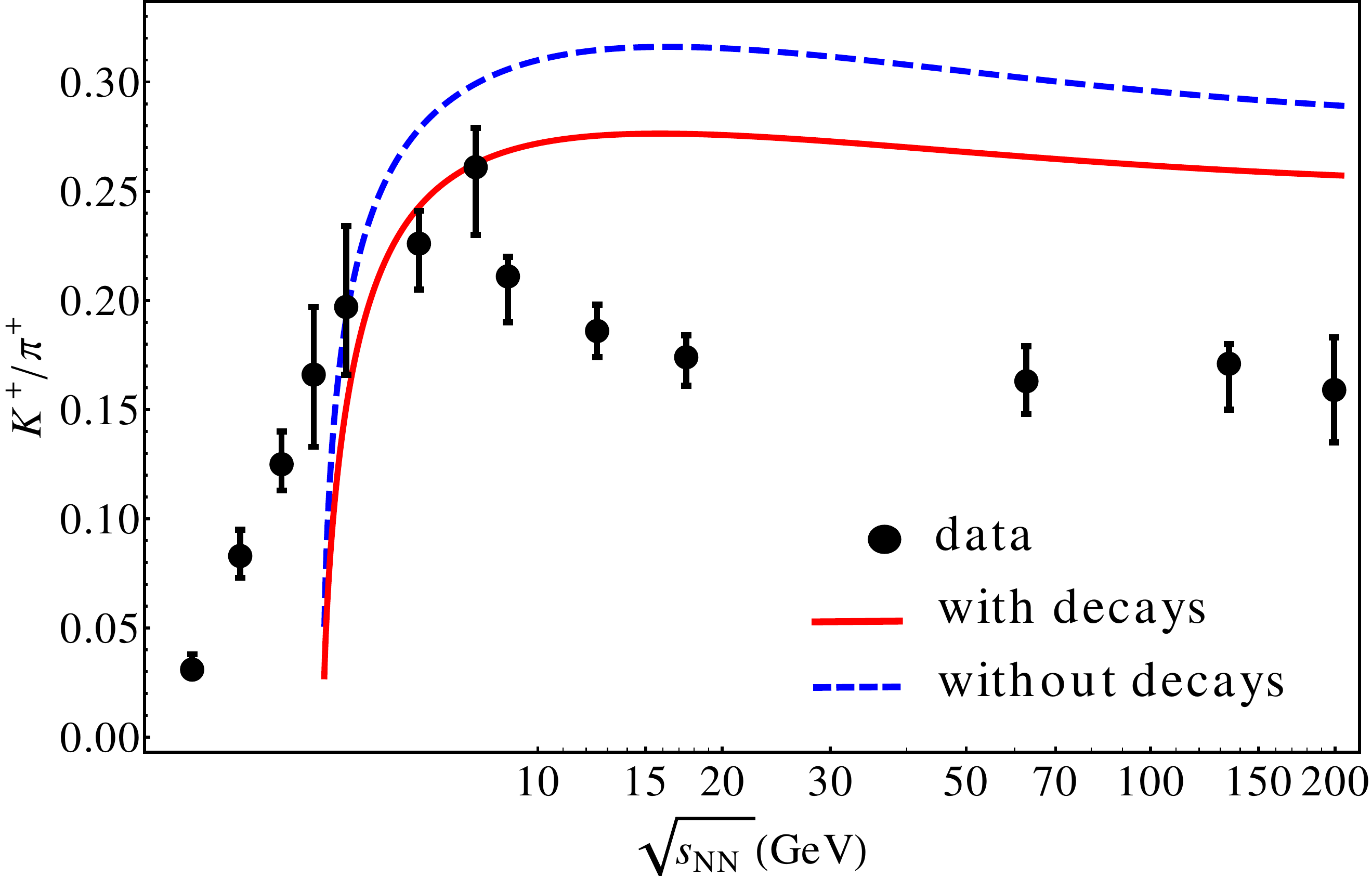}
\end{minipage}
\hfill
\begin{minipage}[b]{0.5\textwidth}
\includegraphics[width=\textwidth]{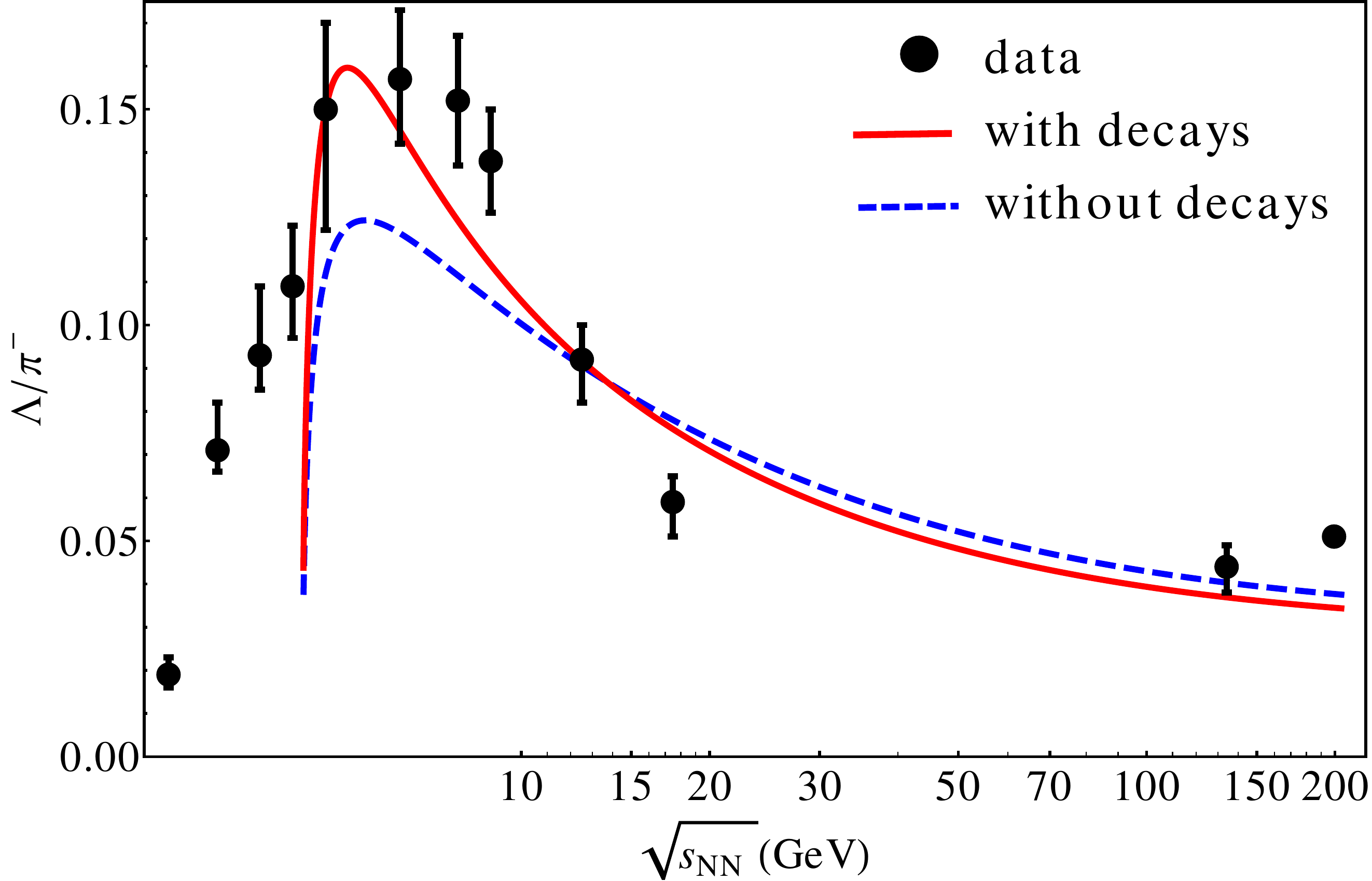}
\end{minipage}
\caption{The ratios $K^+/\pi^+$ (left panel) and $\Lambda/\pi^-$ (right panel) derived with our EOS along the freeze-out curve as a function of $\sqrt{s_{NN}}$; the blue-dashed and red-solid lines refer to the multiplicity ratios without and with the decay contributions respectively; the filled points represents the data (see \cite{Andronic:2009jd} and refences therein).}
\label{fig:kaoni+lambda}
\end{figure}

In a Hadron Resonance Gas model, as the temperature and baryon chemical potential increase new particles are produced, which in turn corresponds to a softening of the EOS. Similarly in our EOS which includes interactions, the effective masses of baryons decrease as functions of the temperature and $\mu_B$ and this again implies a softening of the EOS. This establishes a connection between the two EOSs.
In this approch we provide results solely at the freeze-out stage, whose parametrization curve has been computed with our EOS following the standard procedure \cite{Lavagno:2013iva}.
As we can see in \figurename~\ref{fig:fos}, $\mu_B$ decreases as a function of $\sqrt{s_{NN}}$ while the temperature increases, reaching the saturation value of $T\simeq160\,MeV$ at $\sqrt{s_{NN}}\sim50\,GeV$. 
In \figurename~\ref{fig:regime} we plot the ratio between the density of baryons and the numerical particle density, $n_B/n_{tot}$, and the ratio between the density of mesons and the numerical particle density, $n_M/n_{tot}$, as a function of $\sqrt{s_{NN}}$.
Even with our EOS it is possible to distinguish two regimes of the system at the freeze-out: at very small centre of mass energies ($\sqrt{s_{NN}}\lesssim5\,GeV$) the baryons are the dominating particle species, while at larger energies the mesons are the most abundant particles. The presence of these two regimes and the transition between them are strictly connected with the behaviour of the temperature and the baryon chemical potential as a function of $\sqrt{s_{NN}}$ along the freeze-out curve. At small energies the large $\mu_B$ corresponds to large numerical baryon density, instead at large energies the small values of $\mu_B$ and the moderate value of the temperature lead to a smaller $n_B$ with respect to $n_M$. The change between the baryon-dominated and the meson-dominated regimes takes place in correspondence of the steep increase of $T$ and the strong decrease in $\mu_B$.
These features of the freeze-out curves give rise to the non-monotonic behaviour of the relative number of strange baryons \cite{Cleymans:2004hj}, for example of the ratio $\Lambda/\pi^-$. This ratio should exhibit a sharp peak related to the change between the baryon-dominated and the meson-dominated regimes. Moreover, because of the constraint of the zero-net strangeness and because of the negligible amount of antibaryons, the relative ratios of the strange mesons containing $\bar s$-quark, e.g. $K^+/\pi^+$, ought to follow a similar non-monotonic behaviour with a less pronounced peak. 
In \figurename~\ref{fig:kaoni+lambda} we plot the ratios $K^+/\pi^+$ (left panel), $\Lambda/\pi^-$ (right panel) as a function of $\sqrt{s_{NN}}$ along the freeze-out curve. We consider also the contribution on the particle densities coming from the strong decays of heavier particles included in our hadronic EOS \cite{Satarov:2009zx}.
The decay contributions substantially modify the ratios $K^+/\pi^+$ and $\Lambda/\pi^-$. The inclusion of the decays suppresses the ratio $K^+/\pi^+$ by a factor of nearly $10\%$. The curve associated to $\Lambda/\pi^-$ with the decay contributions is characterised by an higher and sharper maximum with respect to the case in which the decay contributions are not included. Thus, the ratio $\Lambda/\pi^-$ shows the non-monotonic behaviour as a function of $\sqrt{s_{NN}}$, with a peak at small energies $\sqrt{s_{NN}}\sim5\,GeV$, as we expected in a scheme characterised by a transition between a baryon-dominated to a meson-dominated regime.
We notice that the maximum value for $\Lambda/\pi^-\sim0.16$ is comparable to the higher experimental point. Actually, the position of the peak results to be at slightly smaller energies with respect to the data. The $K^+/\pi^+$ ratio obtained is not in good agreement with the data and the peak is not very evident. 
Although our results are not in good agreement with the experimental points, we can conclude that the introduction of the decay contributions allows us to get results which are closer to the data. 
We interpret the better result observed for $\Lambda/\pi^-$ as due to the fact that our model includes only for the baryons their effective masses, while mesons are included as a free gas with only effective chemical potentials. It would be interesting to investigate whether the inclusion of effective masses also for mesons can lead within our model to results closer to the hadron resonance gas ones. In chiral models, for instance, the modification of the meson masses is a natural consequence of the restoration of the chiral symmetry at high energy densities.
Nonetheless, even the hadron resonance gas models reproduce the overall behaviour of the horn, but they do not present a perfect agreement with the data within the whole explored energy range.

\subsection{Roll Over scheme}
\begin{figure}[h!]
\begin{minipage}[b]{0.5\textwidth}
\centering
\includegraphics[width=0.9\textwidth]{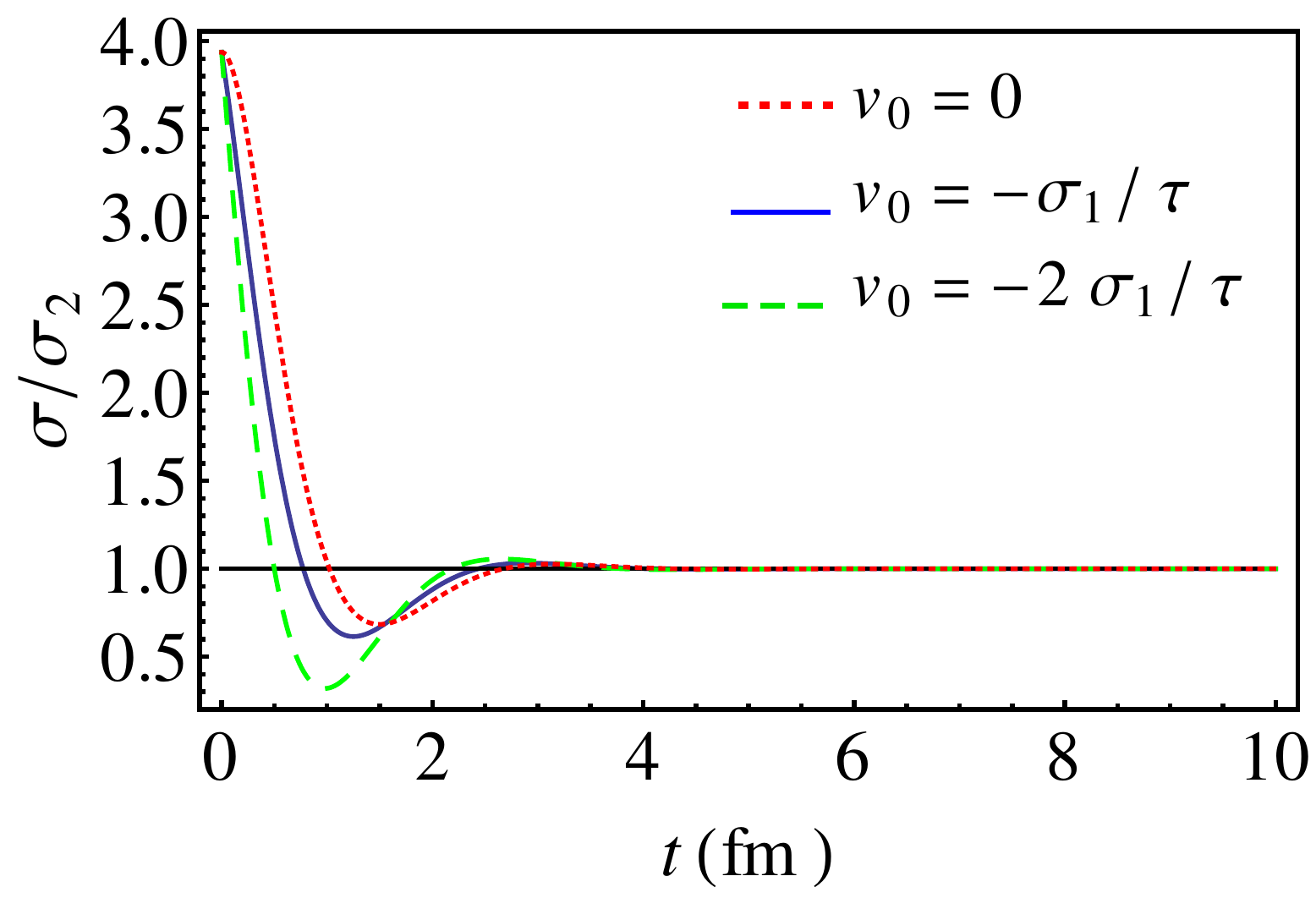}
\end{minipage}
\begin{minipage}[b]{0.5\textwidth}
\centering
\includegraphics[width=\textwidth]{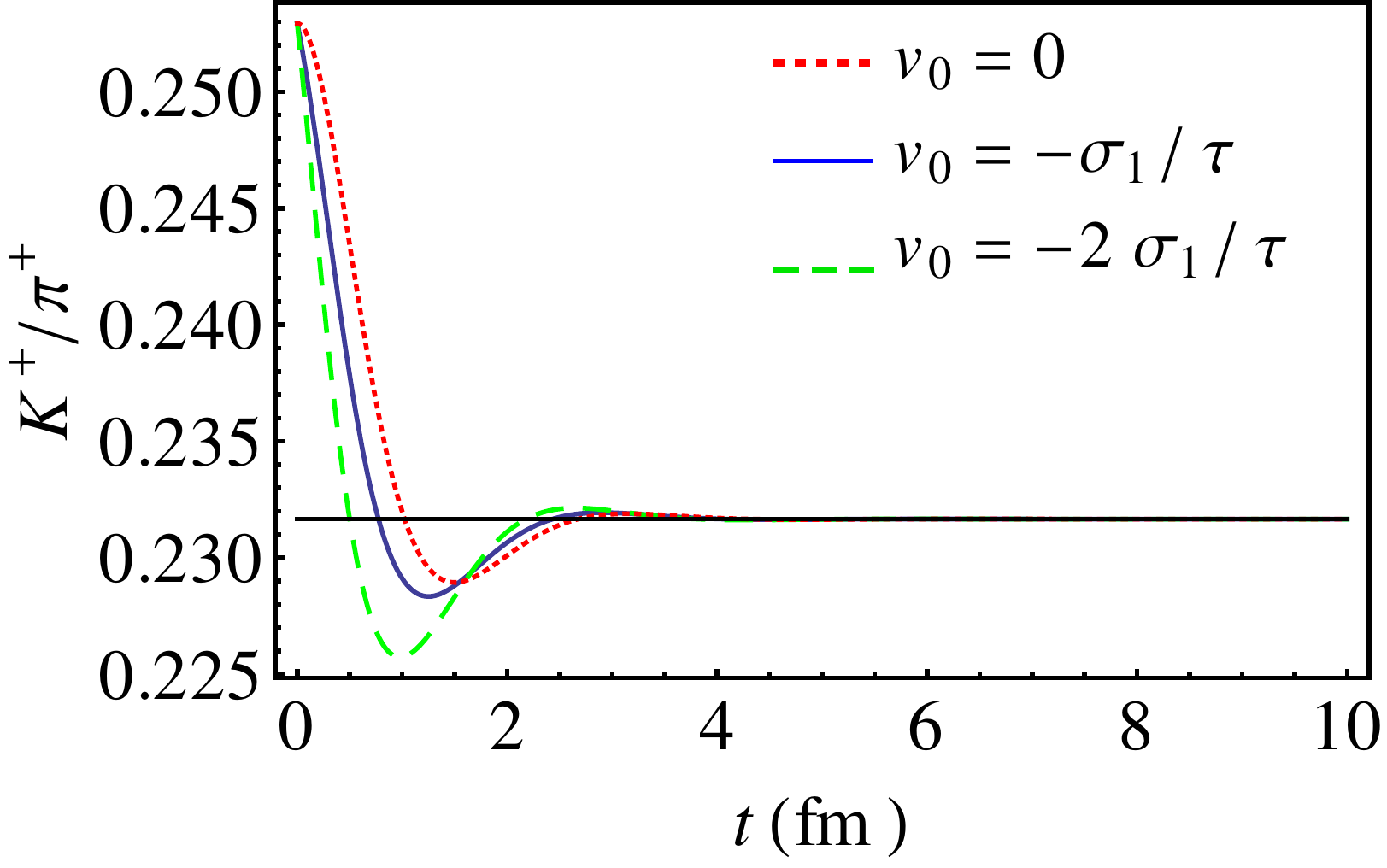}
\end{minipage}
\caption{The ratio $\sigma(t)/\sigma_2$ (left panel) and $K^+/\pi^+$ (right panel) as a function of time; the red-dotted line refers to $v_0=0$; the blue-solid line refers to $v_0=-\sigma_1/\tau$; the green-dashed line refers to $v_0=-2\cdot\sigma_1/\tau$.}
\label{fig:rollmu}
\end{figure}

We solve Eq. \eqref{roll} according to the dynamical scheme in \figurename~\ref{fig:potroll}. The parameters of the system in a pure hadron phase are $\tau=2\,fm$, $\mu_1=750\,MeV$, $\mu_2=550\,MeV$, $T=156\,MeV$ (the temperature does not change significantly and we consider it to be constant). During the expansion the field $\sigma$ develops from $\sigma_1$ to $\sigma_2$, which are the minima of the effective potential in case of $\mu_1$ and $\mu_2$ respectively.
Since the equation of motion \eqref{roll} is a second order differential equation, we need two initial conditions in order to determine a unique solution. Obviously, the condition on the initial position of the field is $\sigma(t=0)=\sigma_1$. Instead, it is not so trivial to fix the initial condition on the first time derivative of the field, $\dot\sigma$. Thus, we investigate different values for the initial velocity $v_0=\dot\sigma=0, -\sigma_1/\tau, -2\cdot\sigma_1/\tau$ and the solution $\sigma(t)$ is shown in \figurename~\ref{fig:rollmu} (left panel).
The fields oscillates around the equilibrium position until $t\sim6\,fm$, time at which the ratio $\sigma(t)/\sigma_2$ becomes equal to $1$ with an error of $\sim10\%$. Due to the friction term the amplitude of the oscillation of $\sigma(t)/\sigma_2$ decreases with time. There are no relevant differences between the three plotted curves associated to different value of $v_0$. The non-equilibrium of the scalar field produces consequences on the particle production and in \figurename~\ref{fig:rollmu} (right panel) we show the effect on the ratio $K^+/\pi^+$. The scheme we have presented is just a simple model, which has to be improved in future. In fact, we know that there are some parameters to be fixed, namely the initial velocity $v_0$ of the field and the expansion rate of the fireball, which enters in the equation of motion as a coefficient of the friction term. The estimate of both parameters is quite uncertain, but this initial velocity actually does not influence the formal result. Moreover, during the expansion of the fireball both $\mu_B$ and $T$ vary, hence we need to apply the scheme to a system evolving from a configuration $(\mu_1,T_1)$ to a configuration $(\mu_2,T_2)$. In conclusion, this scheme would represent a new idea to describe the particle production in heavy-ion collisions, since clearly the described mechanism can modify the particle ratios.

\section{Conclusions}
We have presented a detailed study of the strangeness production in heavy-ion collisions, analysing three alternative interpretations of the horn structure of the $K^+/\pi^+$ ratio: the onset of deconfinement as suggested by the Statistical Model of the Early Stage (SMES), the transition from a baryon-dominated to a meson-dominated hadron gas and the partial chemical equilibrium of the equation of state (EOS) during the fireball expansion. We use for the analysis an EOS, which includes hadron degrees-of-freedom, a hadron-quark mixed phase and quark degrees-of-freedom.
In the confined phase interactions between hadrons are included via meson exchange within a relativistic mean-field model. The baryon chemical potential and baryon density of the system at the early stage, evaluated through the Taub-adiabate procedure, are not negligible and, with the used model for the EOS, we can not explain the horn structure as due to the onset of the deconfinement. On the other hand, using the hadronic EOS we reproduce the energy dependence of the $K^+/\pi^+$ and $\Lambda/\pi^-$ ratios employing the experimental parametrisation of the freeze-out curve. We find that the data on $\Lambda/\pi^-$ can be understood by a transition from a baryon-dominated to a meson-dominated regime. However, the $K^+/\pi^+$ ratio also within this approach could not be reproduced. Finally, we propose a new idea, the roll-over scheme. We find that the scalar meson field $\sigma$ has not reached the minimum of the thermodynamical potential at the end of the hydrodynamical expansion of the fireball. The $\sigma$ plays a crucial role in the particle densities, since the corresponding interaction modifies the effective mass of the hadrons. Consequently the partial equilibrium of the $\sigma$ affects the particle production and hadron abundancies as well.
\\We are aware that statistical models provide only a partial description of the collision process, while transport approches \cite{Cassing:2009vt} are able to describe the whole dynamics and thus achieve more solid predictions on the the strangeness production. Such models may also serve to fix the free parameters of our roll-over scheme.
\acknowledgments
%\section*{Acknowledgment}
A. Palmese gratefully acknowledges A. Lavagno for his crucial and helpful assistance and the 9th International Workshop on CPOD Local Organizing Committee for the opportunity to present this contribution. The authors appreciate the financial support through the "HIC for FAIR" framework of the LOEWE program, HGS HIRe and EMMI.


\begin{thebibliography}{99}
  \bibitem{Gazdzicki:1998vd}
M. Gazdzicki and M.I. Gorenstein,
On the early stage of nucleus-nucleus collisions,
\emph{Acta Phys.Polon.} {\bf B30} (1999) 2705,
[{\tt hep-ph/9803462}].

  \bibitem{Andronic:2009jd}
A. Andronic, P. Braun-Munzinger and J. Stachel,
The Horn, the hadron mass spectrum and the QCD phase diagram: The Statistical model of hadron production in central nucleus-nucleus collisions,
\emph{Nucl.Phys.} {\bf A834} (2010) 237C,
[{\tt nucl-th/0911.4931}].

  \bibitem{Lavagno:2010ah}
A. Lavagno,
Hot and dense hadronic matter in an effective mean field approach,
\emph{Phys.Rev.} {\bf C81} (2010) 044909,
[{\tt nucl-th/1004.0822}].
 

  \bibitem{Muller:1995ji}
H. Muller and B.D. Serot,
Phase transitions in warm, asymmetric nuclear matter,
\emph{Phys.Rev.} {\bf C52} (1995) 2072,
[{\tt nucl-th/9505013}].


  \bibitem{Satarov:2009zx}
L.M. Satarov, M.N. Dmitriev and I.N. Mishustin,
Equation of state of hadron resonance gas and the phase diagram of strongly interacting matter,
\emph{Phys.Atom.Nucl.} {\bf 72} (2009) 1390,
[{\tt hep-ph/0901.1430}].


   \bibitem{Gazdzicki:2010iv}
M. Gazdzicki, M. Gorenstein and P. Seyboth,
Onset of deconfinement in nucleus-nucleus collisions: Review for pedestrians and experts,
\emph{Acta Phys.Polon.} {\bf B42} (2011) 307,
[{\tt hep-ph/1006.1765}].


  \bibitem{Arsene:2006vf}
I.C. Arsene, L.V. Bravina, W. Cassing,Yu.B. Ivanov, A. Larionov and others,
Dynamical phase trajectories for relativistic nuclear collisions,
\emph{Phys.Rev.} {\bf C75} (2007) 034902,
[{\tt nucl-th/0609042}].

  \bibitem{Merdeev:2011bz}
A.V. Merdeev, L.M. Satarov and I.N. Mishustin,
Hydrodynamic modeling of deconfinement phase transition in heavy-ion collisions at NICA-FAIR energies,
\emph{Phys.Rev.} {\bf C84} (2011) 014907,
[{\tt hep-ph/1103.3988}].

  \bibitem{Andronic:2005yp}
A. Andronic, P. Braun-Munzinger and J. Stachel,
\emph{Hadron production in central nucleus-nucleus collisions at chemical freeze-out},
\emph{Nucl.Phys.} {\bf A772} (2006) 167,
[{\tt nucl-th/0511071}].


  \bibitem{Bugaev:2014cha}
K.A. Bugaev, A.I. Ivanytskyi, D.R. Oliinychenko, E.G. Nikonov, V.V. Sagun, and others,
Recent Results of the Hadron Resonance Gas Model and the Chemical Freeze-out of Strange Hadrons,
arXiv:1412.6571v2 (2014),
[{\tt hep-ph/1412.6571}].

  \bibitem{Becattini:2003wp}
F. Becattini, M. Gazdzicki, A. Keranen, J. Manninen, and R. Stock,
\emph{Chemical equilibriumin nucleus nucleus collisions at relativistic energies},
 \emph{Phys.Rev.} {\bf C69} (2004) 024905,
[{\tt hep-ph/0310049}].

  \bibitem{Cleymans:2004hj}
 J. Cleymans, H. Oeschler, K. Redlich,  and S. Wheaton,
Transition from baryonic to mesonic freeze-out,
 \emph{Phys.Lett.} {\bf B615} (2005) 50,
[{\tt hep-ph/0411187}].

  \bibitem{roll}
E. W. Kolb and M. S. Turner,
\emph{The early Universe}
Addison-Wesley Publishing Company 1988.


  \bibitem{Drago:2001gd}
A. Drago, M. Gibilisco, C. Ratti,
Evaporation of the gluon condensate: A Model for pure gauge SU(3)(c) phase transition,
\emph{Nucl.Phys} {\bf 742} (2004) 165,
[{\tt hep-ph/0112282}].

  \bibitem{ivanov}
Yu.B. Ivanov, V.N. Russkikh and V.D. Toneev, ,
Relativistic heavy-ion collisions within 3-fluid hydrodynamics: Hadronic scenario,
\emph{Phys.Rev.} {\bf C73} (2006) 044904,
[{\tt nucl-th/0503088}].


  \bibitem{Lavagno:2013iva}
A. Lavagno, 
Hadronic freeze-out in an effective relativistic mean field model,
\emph{Eur.Phys.J.} {\bf A49} (2013) 102,
[{\tt nucl-th/1311.3094}].

  \bibitem{Cassing:2009vt}
W. Cassing, W. and E.L. Bratkovskaya, 
Parton-Hadron-String Dynamics: an off-shell transport approach for relativistic energies,
\emph{Nucl.Phys.} {\bf A831} (2009) 215,
[{\tt nucl-th/0907.5331}].



\end{thebibliography}
\end{document}